\documentclass[12pt]{article}

\newcommand{\m}{\mathrm}
\newcommand{\be}{\begin{equation}}
\newcommand{\ee}{\end{equation}}
\newcommand{\ba}{\begin{eqnarray}}
\newcommand{\ea}{\end{eqnarray}}

\usepackage{graphicx}
\usepackage{amssymb}
\usepackage{amsmath}
\usepackage[T1]{fontenc} 
\usepackage[ansinew]{inputenc} 
\usepackage[nosort]{cite}
\newcommand{\inbar}{\vrule height1.57ex width.4pt depth0pt}
\newcommand{\SW}{\relax{\hbox{$\ \inbar\kern-.285em{\rm S}$}}}

\begin{document}
\thispagestyle{empty}
\begin{center}

\null \vskip-1truecm \vskip2truecm

{\Large{\bf \textsf{About Magnetic AdS Black Holes}}}

{\Large{\bf \textsf{}}}

{\large{\bf \textsf{}}}

\vskip1truecm

{\large \textsf{Brett McInnes
}}

\vskip0.1truecm

\textsf{\\ National
  University of Singapore}
  \vskip1.2truecm
\textsf{email: matmcinn@nus.edu.sg}\\

\end{center}
\vskip1truecm \centerline{\textsf{ABSTRACT}} \baselineskip=15pt
\medskip

There has recently been a strong revival of interest in quasi-extremal magnetically charged black holes. In the asymptotically flat case, it is possible to choose the magnetic charge of such an object in such a manner that the black hole is surrounded by a corona in which electroweak symmetry is restored on macroscopic scales, a result of very considerable interest. We argue that holographic duality indicates that the asymptotically AdS analogues of these black holes have several interesting properties: the dual theory is only physical if the black hole is required to rotate; in the rotating case, the magnetic field at the poles does not attain its maximum on the event horizon, but rather somewhat outside it; the magnetic field at the equator is not a monotonically decreasing function of the magnetic charge; the electric fields induced by the rotation, while smaller than their magnetic counterparts, are by no means negligible; the maximal electric field often occurs neither at the poles nor at the equator; and so on. Most importantly, in the magnetically charged case it is possible to avoid the superradiant instability to which neutral AdS-Kerr black holes are subject; but the need to avoid this instability imposes upper bounds on the magnetic and electric fields. In some circumstances, therefore, the corona may not exist in the asymptotically AdS case.

\newpage

\addtocounter{section}{1}
\section* {\large{\textsf{1. Magnetic Black Holes in AdS Must Spin}}}
It has long been assumed that the electric charge on astrophysical black holes must be negligible. Recently, however, this belief has been challenged, both for specific observational reasons connected with mergers of black holes \cite{kn:bing}, and for more general reasons based on the sensitive dependence of the radius of the innermost stable circular orbit on the charge \cite{kn:zaj}.

The case of \emph{rotating} charged black holes is particularly interesting, since the usual elementary argument in favour of electrical neutrality can apparently be circumvented in this case. Nevertheless, a very detailed analysis of this question \cite{kn:gong} suggests that even these black holes are probably electrically neutralized (though a marginal loophole, in the case where a magnetosphere is present, remains to be fully closed). For clarity, we will therefore assume henceforth that the electric charge of the black holes to be considered here is zero.

The case of \emph{magnetically} charged black holes \cite{kn:weinberg} has also begun recently to attract renewed attention. Because magnetic monopoles are rare and probably extremely massive, magnetic black holes differ from their electric counterparts: they are not readily neutralized; if they can form at all, they will persist. They are therefore potentially of observational interest. For example, magnetic charge on black holes can affect the merger time of binaries \cite{kn:rong}; again, it is possible, in certain theories, to constrain magnetic charges by studying the ``shadows'' around supermassive black holes \cite{kn:mota}. More dramatically, it has been suggested \cite{kn:juan0,kn:juan1,kn:juan2} that such objects may give rise to enormous magnetic fields, which can restore the full electroweak gauge group in a corona adjacent to the event horizon. All this has led recently to much interest in the phenomenology of these objects: see \cite{kn:yang,kn:ullah} for comprehensive treatments of these developments.

From a theoretical point of view, magnetically charged black holes may lead to a better understanding of the electroweak interaction in a very extreme environment, and similarly of quantum chromodynamics in the form it takes in the presence of extremely strong magnetic fields. (This in itself is a subject which has attracted much interest lately \cite{kn:shubha,kn:swirl}.) One might hope in this way to gain some insight into theories beyond the Standard Model.

It is therefore interesting to embed such black holes in string theory. The natural way to begin such a programme is to consider the asymptotically anti-deSitter versions of these objects.

Magnetically charged asymptotically AdS$_4$ black holes were discussed briefly in \cite{kn:juan0}, where in particular it is pointed out that, by manipulating the magnetic field on the conformal boundary of AdS$_4$, one can arrange for a pair of magnetic black holes to be located at desired positions in the bulk spacetime, with magnetic field lines threading both black holes. This is useful for the study of four-dimensional traversable wormholes, but these objects are clearly of more general interest.

This raises the obvious question: can we exploit the AdS/CFT duality \cite{kn:casa,kn:nat,kn:bag} to constrain magnetically charged asymptotically AdS$_4$ black holes\footnote{For the details of the AdS/CFT correspondence for a magnetically charged four-dimensional bulk black hole, see for example \cite{kn:hartkov}; for the general AdS/CFT correspondence for a four-dimensional bulk, see \cite{kn:ABJM,kn:AdS4}.}? At the very least, we should be able to show that \emph{these black holes are not dual to some system that might be completely unphysical}.

In \cite{kn:juan0,kn:juan1,kn:juan2} it is shown that, due to its Hawking radiation, the magnetic black hole will evolve towards being nearly extremal; precisely because of the existence of the corona, this happens very rapidly, and so near-extremality will be the generic state of the black hole\footnote{See \cite{kn:yc} for the details of the effects of Hawking radiation in such cases; the evolution is not completely straightforward and can proceed in unexpected ways. For example, Hawking radiation can also cause a \emph{rotating} black hole to evolve initially towards extremality \cite{kn:tay1,kn:tay2}, though Censorship is never violated. Note also that astrophysical magnetic black holes can be temporarily perturbed away from extremality by their environments \cite{kn:yang}.}. Such a black hole has a very low Hawking temperature, and hence the dual matter in the asymptotically AdS case is (relatively) cold. Furthermore, the magnetic charge on the black hole induces a non-zero magnetic field $B^{\infty}$ at infinity, so the dual matter is subjected to a substantial magnetic field.

Low-temperature strongly coupled (quark) matter can and probably does exist \cite{kn:annala}, in the cores of certain neutron stars; furthermore, a particular variety of neutron stars, the \emph{magnetars} \cite{kn:magnetar}, often contain matter exposed to very large magnetic fields. Thus it appears that asymptotically AdS$_4$ magnetically charged extremal black holes are holographically dual to matter that is just a more extreme version of a form of matter \emph{which probably does actually exist}, in the cores of certain neutron stars. Thus, it begins to seem that extremal magnetic AdS black holes are dual to a form of matter that is at least not obviously unphysical.

However, an important parameter describing any form of strongly coupled matter is the \emph{chemical potential} $\mu$ \cite{kn:jamal} representing, in this case, the matter density. Cold matter of this kind \emph{always} has a large chemical potential \cite{kn:baym}. In the case we are considering here, we can interpret ``large'' to mean that it is large relative to the maximal value of the other relevant quantity with the same dimension, namely the square root of the magnetic field, $B$. In a magnetar, the field is huge by ordinary standards, typically around $10^{14}$ gauss; but this means $\sqrt{B} \approx 2.6$ MeV, compared to $\mu$ which is estimated \cite{kn:baym} to be in the range of 2 to 6 times nuclear matter density, that is, from 600 MeV to well over 1000 MeV. While we are not suggesting that the dual matter in our case closely resembles the quark matter which may exist in the core of a magnetar, this discussion gives us a guide as to our general expectations: the chemical potential induced at infinity by a magnetic AdS extremal black hole should be at least of the order of magnitude of the square root of the magnetic field at infinity, probably larger.

This is a serious problem here, because the time component of the electromagnetic one-form at infinity for a magnetic AdS-Reissner-Nordstr\"{o}m black hole \emph{vanishes}, and the chemical potential is represented holographically by this component\footnote{See \cite{kn:seam} for a recent discussion of this, in the electric case.}; so the matter dual to such a black hole has zero chemical potential. It seems, then, that magnetic AdS-Reissner-Nordstr\"{o}m extremal black holes are dual, that is, \emph{equivalent}, to an entirely unphysical system (strongly coupled matter with both low temperature and negligible matter density) on the boundary.

We will show that the way to resolve this problem is to make the black hole rotate: that is, to consider magnetic AdS$_4$-Kerr-Newman extremal black holes instead of their AdS$_4$-Reissner-Nordstr\"{o}m counterparts. This induces a non-zero angular momentum density at infinity, but also, more importantly here, a constant but non-zero timelike component in the electromagnetic one-form at infinity; thus we have a non-zero chemical potential there. (It also induces an electric field around the black hole, though not at infinity.) Note that other forms of strongly coupled matter in nature, such as those generated by a generic collision of heavy ions, also typically have high angular momentum densities \cite{kn:littlebang,kn:polar}; so, again, we are considering a more extreme version of a form of matter which does exist. In the case at hand, in view of our discussion above, we will endeavour to use this effect to raise $\mu$ to at least the same order of magnitude as $\sqrt{B^{\infty}}$.

We stress that we are \emph{not} attempting to construct a realistic holographic model of any form of quark matter here. Our objective, rather, is to show that magnetic AdS extremal black holes need not be dual to a flagrantly unphysical system.

Even with this understanding, it is by no means clear that this stratagem can be made to work. It is known \cite{kn:hawkreall} that \emph{all} near-extremal AdS$_4$-Kerr black holes are unstable against the well-known \emph{superradiant instability} \cite{kn:super}. One might suppose that this is also true in the near-extremal AdS$_4$-Kerr-Newman case. If that were so, then adding angular momentum would not solve our original problem.

One objective of this work is to show that these two problems ``solve each other''. Making the purely magnetic extremal black hole rotate can induce a (reasonably large) chemical potential in the theory at infinity, provided that the angular momentum is sufficiently large; in turn, however, a sufficiently large magnetic charge stabilizes the rotating black hole against superradiance. We will give a concrete example of a black hole which embodies both of these virtues.

These results are of course welcome, but clearly they impose potentially interesting bounds on the magnetic charge. Exploring these bounds is another objective of the present work.

We are primarily interested in the \emph{maximal} possible field values (both magnetic and electric) around the black hole. In the non-rotating case, it is clear that the maximal magnetic field occurs on the event horizon, and that, in the extremal case of interest to us here, it is there inversely proportional to the magnetic charge. Leaving aside charge quantization \cite{kn:quant}, then, one can attain arbitrarily large field values simply by taking the charge to be sufficiently \emph{small}; this was exploited in \cite{kn:juan1} to demonstrate the possible existence of coronae around magnetic black holes. In this work, we will investigate the maximal values of the fields in the near-extremal, rotating case.

We find that there are \emph{bounds on the maximal magnetic and electric fields attainable around a rotating purely magnetic AdS$_4$ extremal black hole}, imposed by the requirement that the system be stable against superradiance. In order to explore these bounds, we need to answer two questions.

[1] Since the black hole has to rotate, it is no longer spherically symmetric, and so nor is the magnetic field (or the induced electric field). Thus we need to determine precisely \emph{where} the magnetic and electric fields attain their maxima as functions of position (outside and on the event horizon).

[2] Avoiding superradiance imposes a lower bound on the magnetic charge; this lower bound depends on the angular momentum parameter $a$. We need to understand how the magnetic and electric fields vary with the charge, and to maximize them as a function of the charge (and therefore of $a$).

Combining [1] and [2], we will be able to determine the maximal possible fields for a black hole of this type, given its angular momentum parameter (which is determined holographically by the chemical potential of the field theory at infinity).

These bounds are fixed by the angular momentum parameter of the black hole, by the asymptotic AdS curvature scale $L$, and by the AdS gravitational coupling, or, dually, by the chemical potential, the magnetic field at infinity, and the number of M2-branes in the field theory at infinity. This means that, in some cases, even the maximal magnetic and electric fields may be too small to restore the electroweak symmetry: there may be no corona.

We begin with a brief discussion of the relevant properties of magnetic AdS$_4$-Kerr-Newman black holes.

\addtocounter{section}{1}
\section* {\large{\textsf{2. Magnetic AdS$_4$-Kerr-Newman Black Holes}}}
When the event horizon is topologically spherical, the four-dimensional magnetic AdS-Kerr-Newman metric \cite{kn:cognola} has the form
\begin{flalign}\label{A}
g(\m{AdS}_{a,M,P,L}) = &- {\Delta_r \over \rho^2}\Bigg[\,\m{d}t \; - \; {a \over \Xi}\sin^2\theta \,\m{d}\phi\Bigg]^2\;+\;{\rho^2 \over \Delta_r}\m{d}r^2\;+\;{\rho^2 \over \Delta_{\theta}}\m{d}\theta^2 \\ \notag \,\,\,\,&+\;{\sin^2\theta \,\Delta_{\theta} \over \rho^2}\Bigg[a\,\m{d}t \; - \;{r^2\,+\,a^2 \over \Xi}\,\m{d}\phi\Bigg]^2,
\end{flalign}
where
\begin{eqnarray}\label{B}
\rho^2& = & r^2\,+\,a^2\cos^2\theta, \nonumber\\
\Delta_r & = & (r^2+a^2)\Big(1 + {r^2\over L^2}\Big) - 2Mr + {P^2\over 4\pi},\nonumber\\
\Delta_{\theta}& = & 1 - {a^2\over L^2} \, \cos^2\theta, \nonumber\\
\Xi & = & 1 - {a^2\over L^2}.
\end{eqnarray}
The geometry is characterised by four parameters, all with dimensions of length: $L$ is the asymptotic AdS$_4$ length scale, $a$ is the angular momentum per unit physical mass, and $M$ and $P$ are geometric parameters related \cite{kn:gibperry,kn:komar} to the physical mass $\mathcal{M}$ and the (dimensionless) physical magnetic charge $\mathcal{P}$ by
\begin{equation}\label{C}
\mathcal{M}\;=\;M/(\ell_{\textsf{G}}^2\Xi^2), \;\;\;\;\;\mathcal{P}\;=P/(\ell_{\textsf{G}}\Xi),
\end{equation}
where $\ell_{\textsf{G}}$ is the gravitational length scale in the bulk.

The parameters $a$ and $L$ are related in an interesting way. If we allow $a > L$, then the signature of the metric becomes direction-dependent, changing from $(-, +, +, +)$ to $-, +, -, -)$ for sufficiently small values of $\theta$. Thus the $r$ coordinate becomes timelike (in the ``mostly minus'' signature convention) in some directions, and conformal infinity in those directions is spacelike, as in de Sitter spacetime. While this is interesting, such exotic objects are beyond our scope here, and we will impose the condition
\begin{equation}\label{D}
a \;<\; L.
\end{equation}
Once this is agreed, Cosmic Censorship prohibits values of $a$ arbitrarily close to $L$ if the physical mass and charge of the black hole are fixed: the upper bound on $a/L$ is some number strictly smaller than unity, with a precise value determined by those parameters\footnote{To avoid confusion later, however, we should stress here that, throughout our discussions, we do \emph{not} have any motivation or justification for fixing the mass. In fact, we are primarily interested in quasi-extremal black holes; we will vary the angular momentum and charge independently, and take it that the mass is adjusted so that extremality is maintained.}.

The electromagnetic potential form here is given (see \cite{kn:cognola,kn:aliev,kn:87}) by
\begin{equation}\label{E}
A = \left(-\,{aP\,\cos \theta\over 4\pi \ell_{\textsf{G}} \rho^2}\,+{aP\over 4\pi \ell_{\textsf{G}} \left(r_{\textsf{H}}^2+a^2\right)}\right)\,\m{d}t \;+\;\left({P\,\cos \theta\left(r^2+a^2\right)\over 4\pi \ell_{\textsf{G}}\rho^2} - {P\over 4\pi \ell_{\textsf{G}}}\right)\,\m{d}\phi,
\end{equation}
where the constant terms are fixed by the fact that, in the Euclidean version of the geometry, $\partial_t$ and $\partial_{\phi}$ vanish at the points corresponding to the poles of the Euclidean ``event horizon''; hence the potential one-form, if it is to be regular, must vanish at those points when evaluated on those vectors, and this requires the inclusion of these constant terms. Here $r_{\textsf{H}}$ is the value of the radial coordinate at the event horizon, which is of course related to the other parameters by solving the equation
\begin{equation}\label{F}
\Delta_r(r_{\textsf{H}})\;=\;(r_{\textsf{H}}^2+a^2)\Big(1 + {r_{\textsf{H}}^2\over L^2}\Big) - 2Mr_{\textsf{H}} + {P^2\over 4\pi}\;=\;0
\end{equation}
for its largest root.

Using the orthonormal basis forms with respect to which the fields are radial, one can use (\ref{E}) (and (\ref{C})) to evaluate the magnitude of the  magnetic field outside the event horizon \cite{kn:cognola}:
\begin{equation}\label{G}
B\;=\;{\Xi \, \mathcal{P} \left(r^2-a^2\cos^2\theta\right)\over 4\pi \left(r^2\,+\,a^2\cos^2\theta \right)^2}.
\end{equation}
We note in passing that the presence of the factor $\Xi$ here can be understood as follows: integrating the magnetic field against the element of area in this case, ${r^2+a^2 \over \Xi}\,\sin \theta\,\m{d}\theta\,\m{d}\phi$ (for any surface defined by a fixed $r$), one obtains $\mathcal{P}$, in agreement with Gauss' law, only because of the factor of $\Xi$ on the right side of (\ref{G}). This factor plays a crucial role in our subsequent analysis.

Despite the complete absence of electric charge, the rotation of the magnetically charged black hole induces a non-zero \emph{electric} field, with magnitude given (again by using (\ref{E}) and (\ref{C})) by
\begin{equation}\label{H}
E\;=\;{-\Xi \, \mathcal{P}ra\cos \theta \over 2\pi \left(r^2\,+\,a^2\cos^2\theta\right)^2}.
\end{equation}
This field (which, far from the black hole, falls off in the typical manner of a dipole, that is, as $1/r^3$) will attract or repel nearby electrically charged objects, but of course it cannot be neutralized in that manner and is a permanent feature of the black hole if the magnetic charge does not change. It is a complicated function of the parameters, and will be considered in detail, below.

In order to study the holography of these black holes, we need to understand the electromagnetic fields at conformal infinity. They are obtained by using the same conformal factor\footnote{The electromagnetic field tensor is conformally invariant: therefore its components relative to an \emph{orthonormal} basis are not. Thus one computes the magnetic field at infinity.} as one uses to define the boundary metric. The result is that there is no electric field at infinity, but there \emph{is} a magnetic field at infinity\footnote{It is customary in studies of collisions of heavy ions \cite{kn:umut} to focus on a two-dimensional section through the system, the \emph{reaction plane} or ``$x - z$ plane''. The magnetic field vector is (to a good approximation) parallel to the $y$ - axis perpendicular to the reaction plane, so that only its magnitude plays a role. For magnetic asymptotically AdS black holes, it is useful to think of the magnetic field at infinity in a similar way.}:
\begin{equation}\label{I}
B^{\infty}\;=\;{\Xi \, \mathcal{P}\over 4\pi L^2}.
\end{equation}
Notice that it depends not only on $\mathcal{P}$ but also on $a$, through $\Xi$.

The chemical potential of the strongly coupled matter at infinity is proportional to the timelike component of the electromagnetic one-form at infinity. It is
\begin{equation}\label{J}
\mu\;=\; {3\Xi \, a\mathcal{P}\over 4\pi \left(r_{\textsf{H}}^2+a^2\right)}.
\end{equation}

As we foresaw, this crucial quantity can only be non-zero in the magnetic case if $a \neq 0$: the black hole must spin if the matter at infinity is to be even approximately physical. We regard the chemical potential as being ``given'', though only in the sense that it should not be excessively small ---$\,$ as explained earlier, we are not attempting to set up an even approximately accurate model of realistic quark matter here. As $\mu$ is induced by the black hole angular momentum, we can regard the black hole parameter $a$ as a proxy for it; we adjust $a$ so as to obtain the desired value or range of values for $\mu$. Technically, this means that we must endeavour to express our bounds in terms of quantities determined by $a$.

The Hawking temperature of the black hole is given \cite{kn:cognola} by
\begin{equation}\label{K}
T\;=\;{r_{\textsf{H}} \Big(1\,+\,a^2/L^2\,+\,3r_{\textsf{H}}^2/L^2\,-\,{a^2\,+\,(P^2/4\pi) \over r_{\textsf{H}}^2}\Big)\over 4\pi (a^2\,+\,r_{\textsf{H}}^2)}.
\end{equation}

In order to be more specific, we need to be able to express $r_{\textsf{H}}$ in terms of the other parameters. Unfortunately, equation (\ref{F}) being a quartic equation, this is difficult to do in general.

Fortunately, the problem is much simpler in the extremal case in which we are interested here, because the condition of extremality is equivalent to setting the Hawking temperature equal to zero. The result, from equation (\ref{K}), is
\begin{equation}\label{M}
1\,+\,a^2/L^2\,+\,3r_{\textsf{H}}^2/L^2\,-\,{a^2\,+\,(P^2/4\pi) \over r_{\textsf{H}}^2} \;=\;0,
\end{equation}
which means that
\begin{equation}\label{MM}
r_{\textsf{H}}^2 \;=\; {L^2\over 6}\left(-1 - {a^2\over L^2} + \sqrt{\left(1 + {a^2\over L^2}\right)^2 + {12\over L^2}\left(a^2\,+\,{P^2\over 4\pi}\right)}\right)
\end{equation}
in the extremal case. Using (\ref{C}), we have finally
\begin{equation}\label{N}
r_{\textsf{H}}^2 \;=\; {L^2\over 6}\left(-1 - {a^2\over L^2} + \sqrt{\left(1 + {a^2\over L^2}\right)^2 + {12\over L^2}\left(a^2\,+\,{\ell_{\textsf{G}}^2\left(1 - [a^2/L^2]\right)^2\mathcal{P}^2\over 4\pi}\right)}\right).
\end{equation}

This concludes our discussion of the general properties of magnetic AdS$_4$ black holes. We now turn to the most pressing question for these objects, their stability against superradiance.

\addtocounter{section}{1}
\section* {\large{\textsf{3. Avoiding Superradiance: A Bound on Magnetic Charge }}}
It was observed in \cite{kn:hawkreall} that AdS$_4$-Kerr black holes are at risk of an instability due to superradiance \cite{kn:super}; a sufficient condition for avoiding this is
\begin{equation}\label{L}
\xi \; \equiv \; {aL\,\left(1 + {r_{\textsf{H}}^2\over L^2}\right)\over r_{\textsf{H}}^2 + a^2}\;\leq\;1.
\end{equation}
There is strong reason \cite{kn:reall} to think that this condition is also necessary\footnote{Here, by ``superradiance'', we mean the usual effect due to the existence of an ergosphere. There is another form of black hole ``superradiance'' (see for example \cite{kn:reiss1,kn:reiss2}) due to the effects of charge. In the magnetic case, however, this is most unlikely to be significant, since it would involve \emph{magnetically charged} scalar fields of very low (or zero) mass.}.

It was pointed out in \cite{kn:reall} that \emph{extremal} (and therefore, far worse, sufficiently near-extremal) AdS$_4$-Kerr black holes always violate (\ref{L}). If this were also true of the near-extremal magnetic AdS$_4$-Kerr-Newman black holes with which we are concerned here, then we could not assume that such a black hole can be quasi-stationary, since it would shed its angular momentum to superradiant modes. Thus, since we have argued that holography indicates that these black holes must rotate, it is essential to settle this question.

Using (\ref{N}), we have in the extremal case
\begin{equation}\label{X}
\xi \;= \; {{a\over L}\left(5 - {a^2\over L^2} + \sqrt{\left(1 + {a^2\over L^2}\right)^2 + {12\over L^2}\left(a^2\,+\,\ell_{\textsf{G}}^2\left(1 - [a^2/L^2]\right)^2\mathcal{P}^2/ 4\pi\right)}\right) \over -1 - {5a^2\over L^2} + \sqrt{\left(1 + {a^2\over L^2}\right)^2 + {12\over L^2}\left(a^2\,+\,\ell_{\textsf{G}}^2\left(1 - [a^2/L^2]\right)^2\mathcal{P}^2/ 4\pi\right)}}.
\end{equation}
It is important to note that this expression has a singular limit as $\mathcal{P} \rightarrow 0$: in that limit, this function exceeds unity, for all values of the other parameters (including $a$): this confirms that \emph{all extremal AdS$_4$-Kerr black holes are unstable to superradiance}.

However, if all of the parameters except $\mathcal{P}$ are fixed, and $\xi$ is regarded as a function of $\mathcal{P}$, then one finds that it is a decreasing function; and in all cases, it drops below unity for sufficiently large $\mathcal{P}$. Let us be more explicit about this, using a concrete example.

In order to ensure that the geometry can be self-consistently be treated as (approximately) classical, we need the AdS curvature scale $L$ to be significantly larger than the Planck length; this is a familiar requirement in studies of the AdS/CFT correspondence. To be definite, we set $L = 100$ in Planck units\footnote{When discussing explicit examples, we work with AdS Planck units: that is, we set $\ell_{\textsf{G}} = 1$.}; other choices can of course be made, but that does not affect our general conclusions. Guided by (\ref{D}), we set $a = 50$. These will be our standard choices henceforth, when we need a concrete example.

For this example, the graph of $\xi(\mathcal{P})$ is as shown in Figure 1.
\begin{figure}[!h]
\centering
\includegraphics[width=0.7\textwidth]{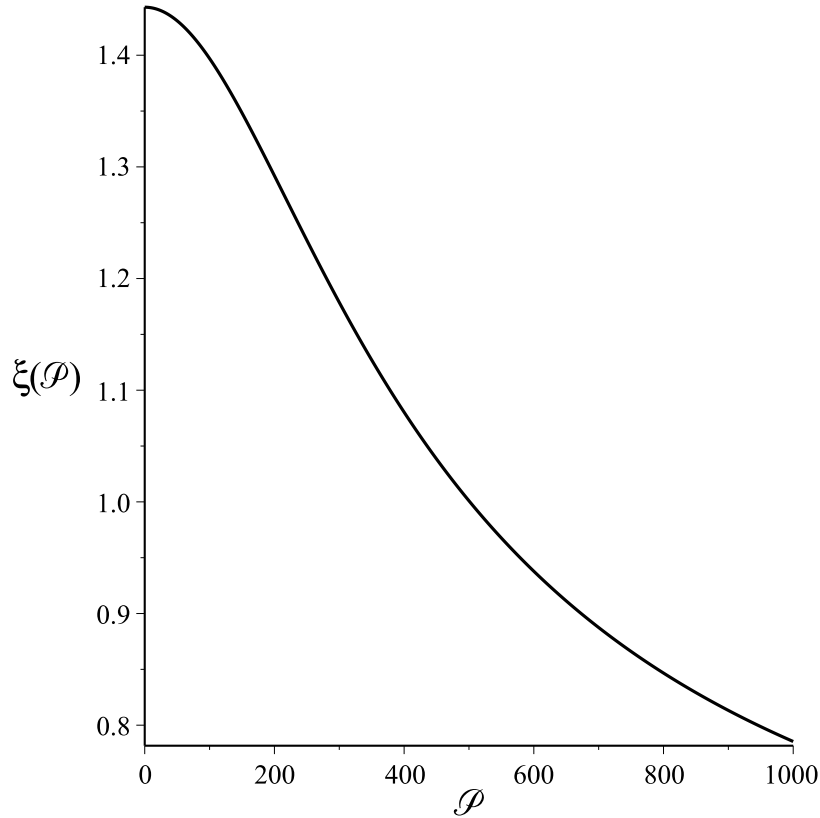}
\caption{$\xi(\mathcal{P})$ with $L = 100, a = 50$. (Units are Planck units.)}
\end{figure}
With these choices of $L$ and $a$, one finds that $\xi \leq 1$ for all values of $\mathcal{P}$ exceeding approximately 501.3.

In short, \emph{the presence of magnetic charge allows us to avoid superradiance}, provided that the magnetic charge is sufficiently large.

The minimal permissible magnetic charge is found by setting $\xi = 1$. After some simplification, one finds that, provided that $a$ is strictly greater than zero (to avoid the singular situation when $\mathcal{P} = 0$), one can avoid a superradiant instability provided that the magnetic charge on the extremal black hole satisfies
\begin{equation}\label{Y}
\mathcal{P} \;\geq\;\mathcal{P}_{\textsf{min}}\;=\;{2\,\sqrt{\pi a L}\over \ell_{\textsf{G}}\left(1-{a\over L}\right)}.
\end{equation}

It turns out that, if one fixes all other parameters and regards the magnetic field as a function of $\mathcal{P}$, then it is \emph{not} a monotonic function; so the consequences of (\ref{Y}) for the magnetic and electric fields are not immediately clear. We will return to that after we determine \emph{where} the maximal fields are to be found.

\addtocounter{section}{1}
\section* {\large{\textsf{4. Where are the Fields Maximal?}}}
We are interested in the case in which the magnitude of the magnetic and electric fields on and outside the event horizon is maximal for a given magnetic charge and angular momentum parameter.

Equation (\ref{G}) shows that the magnetic field is a complicated function of $r$ and $\theta$. In the non-rotating case, it is obvious that this function takes its maximum, on the domain $r \geq r_{\textsf{H}}$, on the event horizon; but it is not clear that this is so in the rotating case. Similar remarks apply to the electric field.

In principle, this is of course a straightforward exercise; but the details are sufficiently intricate that it will be clearest to summarize the results by using our standard example, with $L = 100$ and $a = 50$ in Planck units.

In order to avoid superradiance, we need to take, as explained above, the magnetic charge to be at least 501.3; let us take it to have that value. (As we will see later, this choice of the charge maximizes the magnetic field, all else being equal; we stress again that this is \emph{not} obvious, because the field is not a monotonic function of the charge under these conditions.) For an extremal black hole with these parameters, one finds from (\ref{N}) that $r_{\textsf{H}} \approx 70.7$.

Let us now see what happens at the poles, $\theta = 0$. The graph of $B$ (outside and on the event horizon), which is now a function of $r$ only, is shown as Figure 2.

\begin{figure}[!h]
\centering
\includegraphics[width=0.7\textwidth]{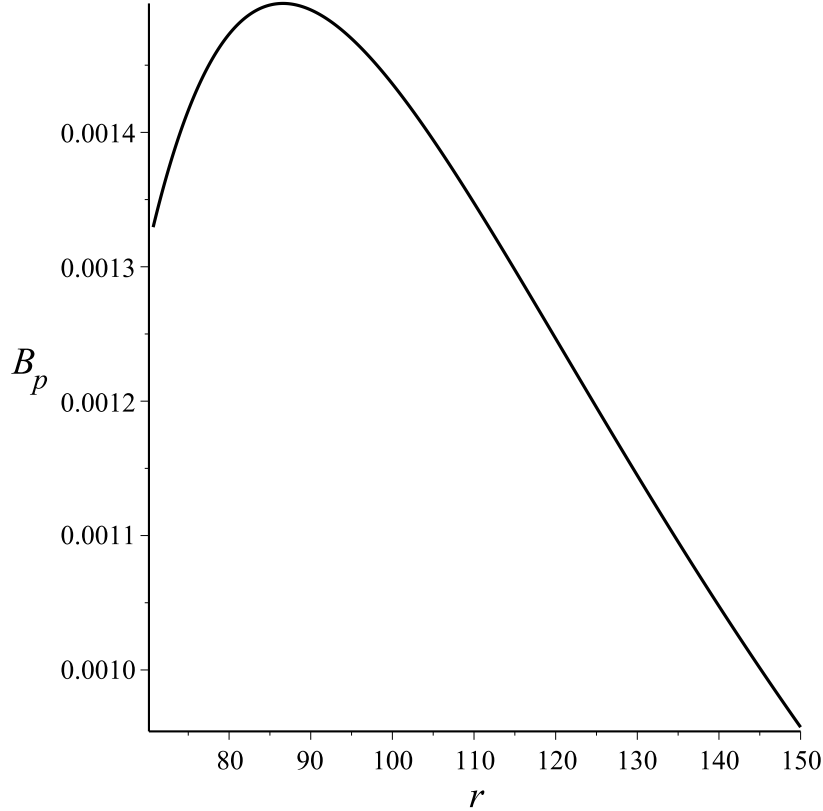}
\caption{The polar magnetic field $B_p$, as a function of the radial coordinate outside the event horizon, for an extremal AdS$_4$-Kerr-Newman magnetic black hole with $L = 100, a = 50, \mathcal{P} = 501.3$. (Units are Planck units.)}
\end{figure}

We see that the field actually attains a local maximum, as a function of $r$, \emph{well outside} the horizon (and therefore, incidentally, outside the ergosphere); it only then decreases towards zero as $r$ increases. This shows the need for caution in the rotating case.

However, a straightforward calculation shows that this maximum, and in fact all of the other local maxima of $B$ as a function of $r$ at intermediate values of $\theta$, is always smaller than the maximal value at the equator, which does occur exactly on the event horizon: see Figure 3.

\begin{figure}[!h]
\centering
\includegraphics[width=0.7\textwidth]{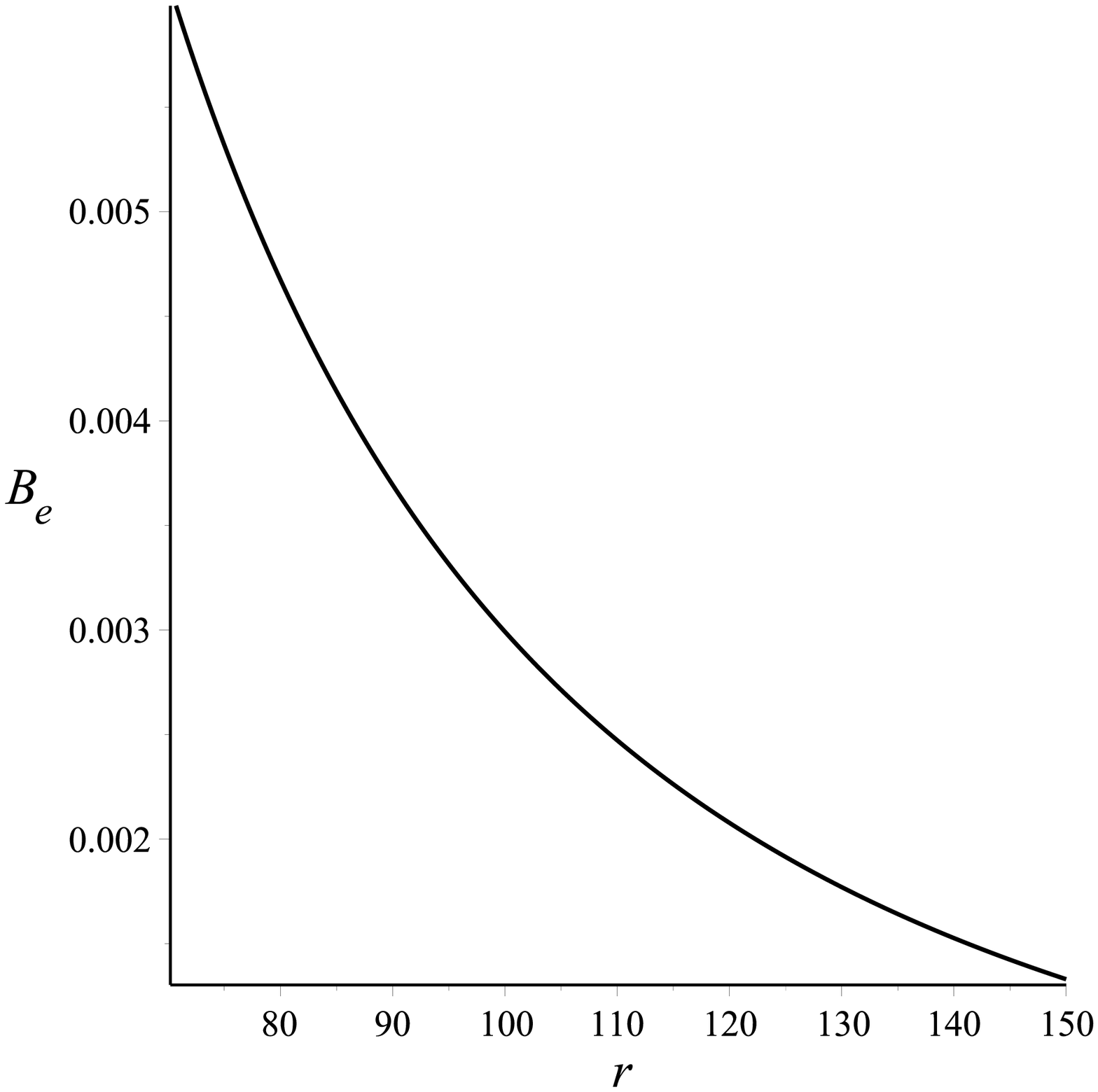}
\caption{The equatorial magnetic field, as a function of the radial coordinate outside the event horizon, for an extremal AdS$_4$-Kerr-Newman magnetic black hole with $L = 100, a = 50, \mathcal{P} = 501.3$. (Units are Planck units.)}
\end{figure}

Unlike the magnetic field, the electric field does always attain its maximum at the event horizon, for each fixed value of $\theta$. However, on the event horizon, it does not always attain its maximum at either the equator or the poles. One finds that the maximum \emph{is} attained at the poles when $a/r_{\textsf{H}} \leq 1/\sqrt{3}$; otherwise, however, the maximum is attained at an intermediate angle $\theta = \theta^+, \; 0 \leq \theta^+ < \pi/2$ (and at the symmetrically located points in the other hemisphere), where
\begin{equation}\label{LL}
\theta^+ \;=\;\arccos\left({r_{\textsf{H}}\over \sqrt{3} a}\right).
\end{equation}
For the parameter values used above, one finds $\theta^+ \approx 0.62$: see Figure 4.

\begin{figure}[!h]
\centering
\includegraphics[width=0.7\textwidth]{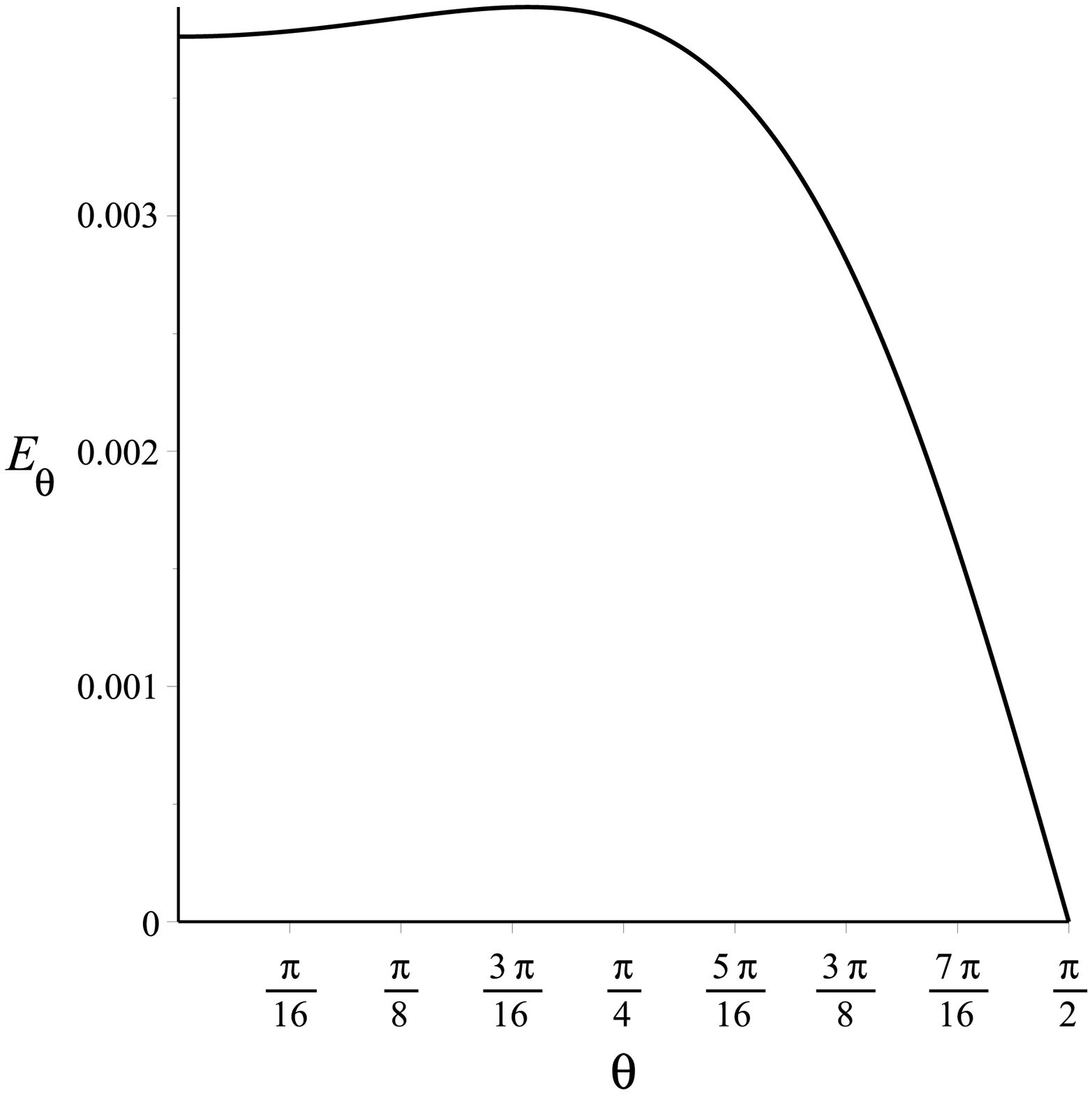}
\caption{The electric field on the event horizon, as a function of $\theta$, for an extremal AdS$_4$-Kerr-Newman magnetic black hole with $L = 100, a = 50, \mathcal{P} = 501.3$. (Units are Planck units.)}
\end{figure}

With these parameter choices, the electric field at $\theta = \theta^+ \approx 0.62$ actually takes a larger value ($\approx 0.0039$ in Planck units) than the magnetic field there ($\approx 0.0022$; this is the value on the event horizon, where it turns out that the magnetic field takes its maximal value as a function of $r$ for this particular value of $\theta$).

In short, the strongest field overall is the magnetic field at the equator, on the event horizon. As we move along the event horizon, away from the equator towards either pole, the magnetic field becomes weaker, but this is partially compensated by the induced electric field, which is distributed in a complex manner that depends on the parameter values (though it is always zero on the equator).

\addtocounter{section}{1}
\section* {\large{\textsf{5. Bounds on the Fields}}}
The magnetic field on the event horizon is, from (\ref{G}),
\begin{equation}\label{O}
B(r = r_{\textsf{H}})\;=\;{\left(1 - [a^2/L^2]\right) \, \mathcal{P} \left(r_{\textsf{H}}^2-a^2\cos^2\theta\right)\over 4\pi \left(r_{\textsf{H}}^2\,+\,a^2\cos^2\theta \right)^2}.
\end{equation}

We saw earlier that the global maximum for the magnetic field, for fixed $\mathcal{P}$ and $a$, occurs on the horizon, at the equator; so this maximum is
\begin{equation}\label{P}
B^+\;=\;{\left(1 - [a^2/L^2]\right) \, \mathcal{P} \over 4\pi r_{\textsf{H}}^2}.
\end{equation}
That is, $B^+$ is the maximal magnetic field, when the latter is regarded as a function of position (\emph{not} of the parameters $a$ and $\mathcal{P}$).

From equation (\ref{N}) (and using (\ref{C})) we have now, in the extremal case,
\begin{equation}\label{Q}
B^+\;=\;{3 \left(1 - [a^2/L^2]\right) \, \mathcal{P} \over 2\pi L^2\left(-1 - {a^2\over L^2} + \sqrt{\left(1 + {a^2\over L^2}\right)^2 + {12\over L^2}\left(a^2\,+\,\ell_{\textsf{G}}^2\left(1 - [a^2/L^2]\right)^2\mathcal{P}^2/ 4\pi\right)}\right)}.
\end{equation}

As an application of these explicit formulae, we now ask: how does the maximal magnetic field vary with the magnetic charge?

Let us recall first the answers to this question for extremal \emph{asymptotically flat} four-dimensional magnetic black holes, and for extremal asymptotically AdS$_4$ but non-rotating magnetic black holes.

In the asymptotically flat, non-rotating case, one has simply $r_{\textsf{H}} = \mathcal{P}\ell_{\textsf{G}}/\sqrt{4\pi}$; so the area of the event horizon becomes \emph{smaller} as the magnetic charge is reduced while maintaining extremality\footnote{Of course, this entails a corresponding reduction of the mass, so this does not contradict the familiar dictum that increasing the charge on a black hole of fixed mass reduces the entropy.}. In that case the maximal magnetic field at the event horizon is just
\begin{equation}\label{UU}
B^+_{\textsf{AF}}\;=\;{1\over \mathcal{P}\ell_{\textsf{G}}^2},
\end{equation}
that is, it is \emph{large} for small magnetic charge, as is stressed in \cite{kn:juan1}. (This is reminiscent of the fact that the spacetime curvature near the horizon of a Schwarzschild black hole is largest for black holes with the smallest masses.) Indeed, leaving aside the constraint imposed by charge quantization, this means that the magnetic field at the event horizon can be made \emph{arbitrarily} large by choosing a sufficiently small $\mathcal{P}$. Thus it is possible to attain such large magnetic fields that the electroweak symmetry can be restored in a corona around the event horizon.

This is still true even in the asymptotically AdS$_4$, non-rotating case: setting $a = 0$ in (\ref{Q}), we have
\begin{equation}\label{V}
B^+(a = 0)\;=\;{3\mathcal{P} \over 2\pi L^2\left(-1 + \sqrt{1 +  {3\ell_{\textsf{G}}^2\mathcal{P}^2/ \pi L^2}}\right)}\;=\;{1\over \mathcal{P}\ell_{\textsf{G}}^2} + {3\mathcal{P}\over 4\pi L^2} + \mathcal{O}(\mathcal{P}^3),
\end{equation}
which is of course unbounded above as $\mathcal{P}$ is reduced.

However, the analogous statement is \emph{false} when the magnetic AdS$_4$ black hole rotates, no matter how small the angular momentum may be: from equation (\ref{Q}) we have
\begin{equation}\label{W}
B^+\;=\;{3\left( 1 - {a^2\over L^2}\right)\mathcal{P}\over 2\pi L^2\left(-1 - {a^2\over L^2} + \sqrt{1 + 14{a^2\over L^2} + {a^4\over L^4}}\right)} + \mathcal{O}(\mathcal{P}^3);
\end{equation}
that is, there is no term in $\mathcal{P}^{-1}$. This means that it is no longer the case that the fields are maximized by choosing $\mathcal{P}$ to be arbitrarily small. Instead, the magnetic field is small for very small charges, and then \emph{increases} with the charge until it reaches a maximum: only for still larger charges does it begin to decrease, as shown in Figure 5 for a typical choice of parameters.
\begin{figure}[!h]
\centering
\includegraphics[width=0.7\textwidth]{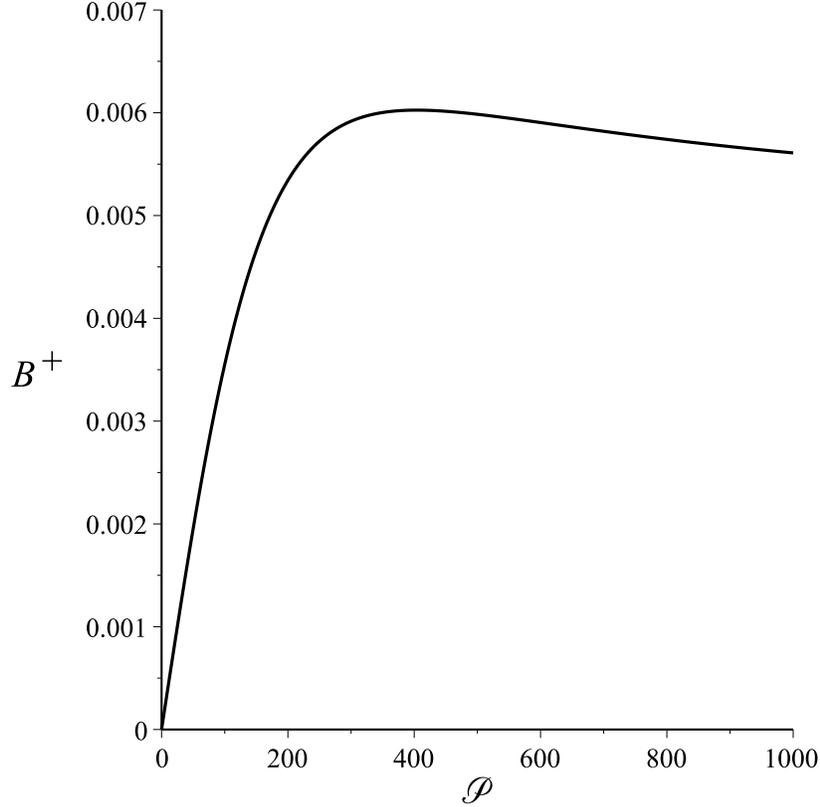}
\caption{The magnetic field, evaluated on the event horizon at the equator, as a function of the physical charge, for an extremal AdS$_4$-Kerr-Newman magnetic black hole with $L = 100, a = 50$. (Units are Planck units.)}
\end{figure}

In short, for a given value of the angular momentum per unit mass, the maximal magnetic field can either increase or decrease if the magnetic charge is increased. This is important when we take into account the condition for superradiance to be avoided, to which we now turn.

A lengthy but elementary computation shows that $\mathcal{P} \;=\;\mathcal{P}_{\textsf{min}}$ always occurs on the \emph{decreasing} branch of $B^+$ regarded as a function of $\mathcal{P}$. It follows that the maximal magnetic field that can be attained by a \emph{stable} extremal AdS$_4$-Kerr-Newman black hole is given by its value on the equator of the event horizon, when the magnetic charge is precisely $\mathcal{P}_{\textsf{min}}$, as given in equation (\ref{Y}). That is, the maximal possible magnetic field for a given value of $a$ (taking into account both variations with position \emph{and} with the magnetic charge, and therefore denoted by $B^{++}$) is given by
\begin{equation}\label{Z}
B^{++}\;=\;{3 \left(1 + [a/L]\right) \, \sqrt{\pi a L} \over \pi \ell_{\textsf{G}} L^2\left(-1 - {a^2\over L^2} + \sqrt{\left(1 + {a^2\over L^2}\right)^2 + {12\over L^2}\left(a^2\,+\,\left(1 + [a/L]\right)^2 a L\right)}\right)}.
\end{equation}

To summarize: the expression in (\ref{Z}) is the upper bound on the possible magnetic fields around a stable extremal magnetic AdS$_4$-Kerr-Newman black hole of given angular momentum to mass ratio $a$. With fixed $L$, it is progressively more stringent for increasing $a$. (It is not, however, arbitrarily restrictive: the limiting value of $B^{++}$ as $a \rightarrow L$ is just $1/\left(\sqrt{\pi}\ell_{\textsf{G}}L\right)$.)

The situation regarding the electric field is similar, though slightly more technically involved.

The electric field also attains is maximum, for given $\mathcal{P}$ and $a$ and for points with $r \geq r_{\textsf{H}}$, on the horizon. There, from (\ref{H}), it is given by
\begin{equation}\label{R}
E(r = r_{\textsf{H}})\;=\;{-\Xi \, \mathcal{P}r_{\textsf{H}}a\cos \theta \over 2\pi \left(r_{\textsf{H}}^2\,+\,a^2\cos^2\theta\right)^2}.
\end{equation}

As we saw earlier, the electric field on the event horizon attains its maximum either at the poles, if $a$ is small relative to $r_{\textsf{H}}$, or at an intermediate value of $\theta$, $\theta^+$; the maximal value of the electric field magnitude for given values of $a$ and $\mathcal{P}$ is then either
\begin{equation}\label{S}
E^+\left(a/r_{\textsf{H}} \leq 1/\sqrt{3}\right)\;=\;{- 3 \sqrt{6} \,\Xi \, \mathcal{P} a \,\sqrt{-1 - {a^2\over L^2} + \sqrt{\left(1 + {a^2\over L^2}\right)^2 + {12\over L^2}\left(a^2\,+\,{\ell_{\textsf{G}}^2\left(1 - [a^2/L^2]\right)^2\mathcal{P}^2\over 4\pi}\right)}}\over \pi \,L^3\left(-1\,+\,{5a^2\over L^2}\,+\,\sqrt{\left(1 + {a^2\over L^2}\right)^2 + {12\over L^2}\left(a^2\,+\,{\ell_{\textsf{G}}^2\left(1 - [a^2/L^2]\right)^2\mathcal{P}^2\over 4\pi}\right)}\right)^2}
\end{equation}
or
\begin{equation}\label{T}
E^+\left(a/r_{\textsf{H}} \geq 1/\sqrt{3}\right)\;=\;{- 9 \sqrt{3}\, \Xi \,\mathcal{P}\over 16 \pi L^2 \left(-1 - {a^2\over L^2} + \sqrt{\left(1 + {a^2\over L^2}\right)^2 + {12\over L^2}\left(a^2\,+\,{\ell_{\textsf{G}}^2\left(1 - [a^2/L^2]\right)^2\mathcal{P}^2\over 4\pi}\right)}\right)},
\end{equation}
where (\ref{LL}) has been used.

One can show that, for each given point on the horizon, the electric field behaves, as the charge is varied, in much the same way as the magnetic field: that is, it increases for small charges, but decreases for larger charges. As in the magnetic case, the field decreases over the range of charges permitted by the condition that superradiance be avoided, so the maximal fields are always attained at the minimal permitted charge, as given in equation (\ref{Y}).

One can also show that, when $\mathcal{P} = \mathcal{P}_{\textsf{min}}$, the condition $a/r_{\textsf{H}} \leq, \, \geq 1/\sqrt{3}$ can be expressed more directly as $a/L \leq, \, \geq 1/3$. Thus the overall maximal possible electric fields (that is, for all locations outside the black hole and for all values of the charge permitted by the need to avoid superradiance) for a given value of $a$ are as follows: first, if $a/L \leq 1/3,$ then
\begin{equation}\label{SS}
E^{++}\left(a/L \, \leq \,1/3\right)\;=\;{- 3 \sqrt{6} \,\Xi \, \mathcal{P}_{\textsf{min}} a \,\sqrt{-1 - {a^2\over L^2} + \sqrt{\left(1 + {a^2\over L^2}\right)^2 + {12\over L^2}\left(a^2\,+\,{\ell_{\textsf{G}}^2\left(1 - [a^2/L^2]\right)^2\mathcal{P}_{\textsf{min}}^2\over 4\pi}\right)}}\over \pi \,L^3\left(-1\,+\,{5a^2\over L^2}\,+\,\sqrt{\left(1 + {a^2\over L^2}\right)^2 + {12\over L^2}\left(a^2\,+\,{\ell_{\textsf{G}}^2\left(1 - [a^2/L^2]\right)^2\mathcal{P}_{\textsf{min}}^2\over 4\pi}\right)}\right)^2};
\end{equation}
in this case, the maximal field occurs at the poles. Secondly, if $a/L \geq 1/3,$ the maximal field is
\begin{equation}\label{TT}
E^{++}\left(a/L \, \geq \,1/3\right)\;=\;{- 9 \sqrt{3}\, \Xi \,\mathcal{P}_{\textsf{min}}\over 16 \pi L^2 \left(-1 - {a^2\over L^2} + \sqrt{\left(1 + {a^2\over L^2}\right)^2 + {12\over L^2}\left(a^2\,+\,{\ell_{\textsf{G}}^2\left(1 - [a^2/L^2]\right)^2\mathcal{P}_{\textsf{min}}^2\over 4\pi}\right)}\right)};
\end{equation}
but now the maximum occurs at $\theta = \theta^+$, with a value that may be read off from Figure 6.
\begin{figure}[!h]
\centering
\includegraphics[width=0.7\textwidth]{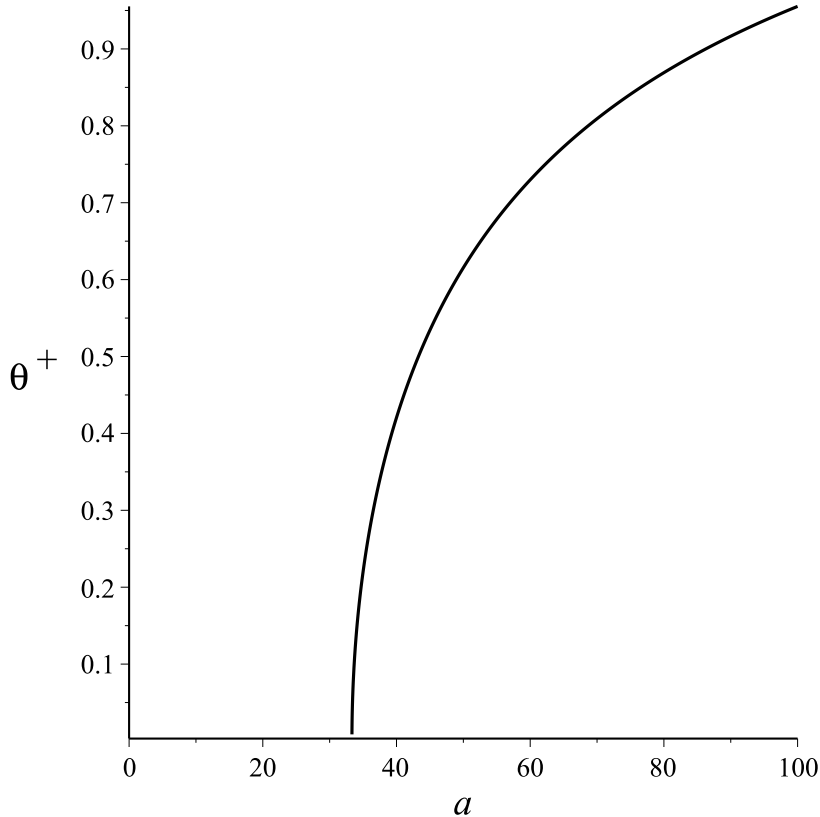}
\caption{The value of $\theta$ which gives the position, on the event horizon, where the induced electric field is maximal, for an extremal AdS$_4$-Kerr-Newman magnetic black hole with $L = 100,$ as a function of $a$. (Units are Planck units; bear in mind that, for $a/L < 1/3$, the maximal field occurs at $\theta = 0$.)}
\end{figure}

In both equations, $\mathcal{P}_{\textsf{min}}$ is to be regarded as a function of $a$, through equation (\ref{Y}). As in the case of the magnetic field, these bounds become steadily more restrictive for larger $a$. Again, however, they are not arbitrarily restrictive: the limiting value of $E^{++}$ as $a \rightarrow L$ is just $1/\left(2\sqrt{\pi}\ell_{\textsf{G}}L\right)$, exactly half of the corresponding quantity in the magnetic case.

In the specific example we have been considering, with $a = 50,$ $L = 100$ (in Planck units), we find that the overall maximum possible fields are (in natural rather than Planck units)
\begin{equation}\label{ALPHA}
B^{++}\left(a = 50, L = 100\right)\;\approx \; 0.00598/\ell_{\textsf{G}}^2; \;\;\;\; E^{++}\left(a = 50, L = 100\right)\;\approx \;0.00389/\ell_{\textsf{G}}^2,
\end{equation}
where the magnetic field maximum occurs along the equator, and its electric counterpart occurs at $\theta \approx 0.62$ (about 36 degrees).

As to the value of $\ell_{\textsf{G}}$, it is (in principle) determined holographically through the ``dictionary'', which in this case states that $\ell_{\textsf{G}}^2 = 3L^2/(2N_{\textsf{c}})^{3/2}$, where $N_{\textsf{c}}$ refers to the presence of this number of coincident M2-branes in the dual field theory \cite{kn:ABJM,kn:AdS4}. In the present case, $N_{\textsf{c}} \approx 483$, which is indeed ``large'', as usual in discussions of the AdS/CFT correspondence.

Our original observation was that the black hole must rotate in order that the chemical potential describing the boundary theory should be substantial, as it is for any form of cold strongly coupled matter. Let us now confirm that this is the case.

\addtocounter{section}{1}
\section* {\large{\textsf{6. The Chemical Potential}}}
We can now compute the chemical potential for matter dual to an extremal magnetic black hole, with magnetic charge chosen, as above, to maximize the fields subject to the condition of stability against superradiance. The result is, from (\ref{J}) and (\ref{N}),
\begin{equation}\label{U}
\mu\;=\; {9\left(1 - [a^2/L^2]\right) \,a\, \mathcal{P}_{\textsf{min}} \over 2\pi L^2 \left(-1\,+\,{5a^2\over L^2}\,+\,\sqrt{\left(1 + {a^2\over L^2}\right)^2 + {12\over L^2}\left(a^2\,+\,\ell_{\textsf{G}}^2\left(1 - [a^2/L^2]\right)^2\mathcal{P}_{\textsf{min}}^2/ 4\pi\right)}\right)};
\end{equation}
as above, $\mathcal{P}_{\textsf{min}}$ is regarded as a function of $a$, using equation (\ref{Y}).

The magnetic field at infinity in this case is, from (\ref{I}) and (\ref{N}),
\begin{equation}\label{BETA}
B^{\infty}\;=\;{\left(1 - [a^2/L^2]\right)\, \mathcal{P}_{\textsf{min}}\over 4\pi L^2};
\end{equation}
again, (\ref{Y}) allows us to regard this as a function of $a$.

As explained earlier, we measure the size of the chemical potential by computing the dimensionless quantity $\mu/\sqrt{B^{\infty}}$. It turns out that this is a monotonically increasing function of $a$, which very quickly rises to values well over unity; however, it is bounded, with limiting value as $a \rightarrow L$ given as follows:
\begin{equation}\label{GAMMA}
{\mu\over \sqrt{B^{\infty}}}\;<\;{3\,\sqrt{L}\over 2\,\pi^{1/4}\,\sqrt{\ell_{\textsf{G}}}}.
\end{equation}
For typical values of $a$, the function is close to its upper bound: see Figure 7 for the case of $L = 100$ in Planck units. (For the case we have been taking in examples, $a = 50$, it is equal to $\approx 10.9$.)

\begin{figure}[!h]
\centering
\includegraphics[width=0.7\textwidth]{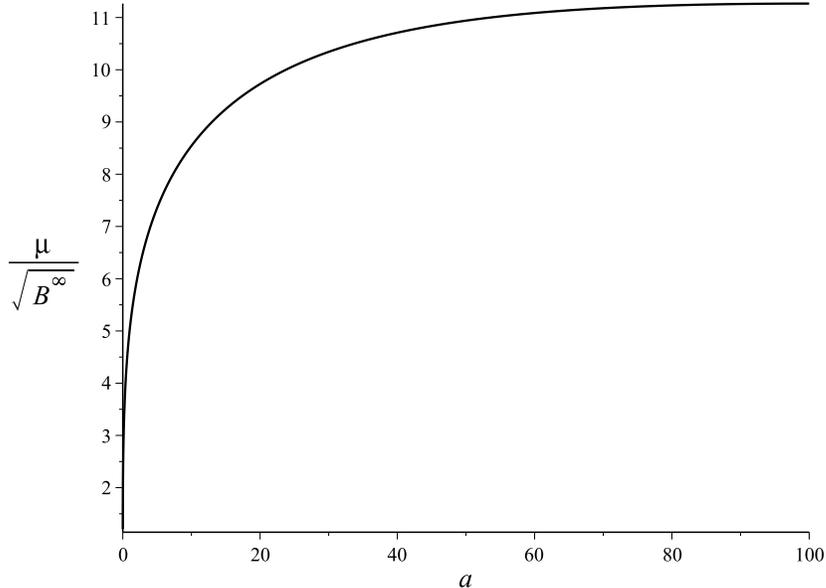}
\caption{$\mu/\sqrt{B^{\infty}}$ as a function of $a$, for $L = 100$ (Planck units). The upper bound is $\approx 11.3$ in this case.}
\end{figure}

If $\mu/\sqrt{B^{\infty}}$ is given, then we can use equations (\ref{U}) and (\ref{BETA}) to compute $a$ (given $L$ and $\ell_{\textsf{G}}$). We can then compute the maximal possible magnetic and electric fields using equations (\ref{Z}) and either (\ref{SS}) or (\ref{TT}).

We conclude from (\ref{GAMMA}) that, since $L$ must always be substantially larger than $\ell_{\textsf{G}}$ if the spacetime is to be semi-classical, then $\mu$ is indeed reasonably large relative to $\sqrt{B^{\infty}}$ for all but an extremely narrow range of values of $a$. In short, absent severe fine-tuning, making the magnetic extremal black hole rotate does induce an acceptable chemical potential at infinity.

\addtocounter{section}{1}
\section* {\large{\textsf{7. Conclusion}}}
We have used the AdS/CFT duality as a guide to the physics of asymptotically AdS$_4$ magnetically charged black holes. However, we have only used it in the most basic possible way: all we required was that the boundary field theory should not be obviously unacceptable. Even this minimal requirement has nevertheless yielded interesting constraints on these objects. For example, we have learned that they can be stable and that the magnetic and electric fields around them cannot be arbitrarily prescribed; so that the coronae studied in \cite{kn:juan1} may not always be possible in the AdS case.

In fact, the field theory at infinity in this case is somewhat exotic: it is the infrared fixed point of a maximally supersymmetric SU(N) gauge theory at large N, describing the low-energy behaviour of M2 branes. Nevertheless the hope has been expressed \cite{kn:herzog} that such field theories might have much to teach us regarding general features of strongly coupled CFTs of the kind that appear in studies of critical phenomena. Perhaps, then, it might be possible to use the AdS/CFT correspondence to deduce further constraints on magnetic AdS$_4$ black holes, again by insisting that the boundary theory should be physically reasonable in some specific sense.

\addtocounter{section}{1}
\section*{\large{\textsf{Acknowledgements}}}
The author thanks Dr Soon Wanmei for valuable discussions.

\end{document}